\documentclass[reprint,aps,prl,twocolumn,notitlepage,nofootinbib,
showpacs,preprintnumbers,superscriptaddress]{revtex4-1}
\usepackage[T1]{fontenc}
\usepackage[utf8]{inputenc}
\usepackage{bm}
\usepackage{amsmath}
\usepackage{amssymb}
\usepackage{esint}
\usepackage{color}
\usepackage{graphicx}
\PassOptionsToPackage{normalem}{ulem}
\usepackage{ulem}

\usepackage{times}
\usepackage{hyperref}
\usepackage{amsthm,bm,mathrsfs,stmaryrd,amsmath}

\newcommand{\be}{\begin{equation}}
\newcommand{\ee}{\end{equation}}
\newcommand{\ba}{\begin{eqnarray}}
\newcommand{\ea}{\end{eqnarray}}

\newcommand{\nn}{\nonumber\\}

\def\pa{\partial}

\def\G{\Gamma}
\def\d{\delta}
\def\D{\Delta}

\def\t{\tau}

\def\O{\Omega}

\def\be{\begin{eqnarray}}
\def\ee{\end{eqnarray}}
\def\D{\Delta}
\def\t{\tau}

\def\D{\Delta}

\def\G{\Gamma}
\def\d{\delta}
\def\nn{\nonumber\\}
\def\pa{\partial}

\newcommand\<\langle
\renewcommand\>\rangle

\begin{document}

\title{Taming Dyson-Schwinger Equations with Null States}

\author{Wenliang Li}
\email{liwliang3@mail.sysu.edu.cn}
\affiliation{School of Physics, Sun Yat-Sen University, Guangzhou 510275, China}

\begin{abstract}
In quantum field theory, 
the Dyson-Schwinger equations are an infinite set of coupled equations 
relating $n$-point Green's functions in a self-consistent manner.  
They have found important applications in nonperturbative studies, ranging from 
quantum chromodynamics and hadron physics to strongly correlated electron systems. 
However, they are notoriously formidable to solve. 
One of the main obstacles is that a finite truncation of the infinite system is underdetermined. 
Recently,  Bender {\it et al.} [Phys. Rev. Lett. 130, 101602 (2023)] proposed to make use of the large-$n$ asymptotic behaviors 
and successfully obtained accurate results in $D=0$ spacetime.  
At higher $D$, it seems more difficult to deduce the large-$n$ behaviors. 
In this Letter, we propose another avenue in light of the null bootstrap. 
The underdetermined system is solved by imposing the null state condition. 
This approach can be extended to $D>0$ more readily. 
As concrete examples, we show that the cases of $D=0$ and $D=1$ indeed converge to the exact results 
for several Hermitian and non-Hermitian theories of the $g\phi^n$ type, including the complex solutions.  
\end{abstract}

\maketitle 
\paragraph{Introduction.} 
In the early stages of quantum field theory, 
the Dyson-Schwinger (DS) equations \cite{Dyson:1949ha,Schwinger:1951ex,Schwinger:1951hq} 
were formulated as an alternative to operator theory. 
They furnish nonperturbative self-consistency relations for the $n$-point Green's functions.  
To make concrete predictions, one usually needs to restrict to a finite subset of the DS equations 
for low-point Green's functions. 
However, this is known to be an {\it underdetermined} system,  
as higher DS equations involve higher-point Green's functions \cite{Bender:1988bp}. 
One needs to introduce additional constraints to solve the DS system. 

A simple scheme is to close the system by setting high-point connected Green's functions to zero. 
However, as emphasized recently in \cite{Bender:2022eze}, the results from this naive procedure do not converge to the exact values. 
The resolution proposed in \cite{Bender:2022eze} is to make use of the asymptotic behaviors of 
the connected Green's functions at large $n$. 
This has been successfully carried out at $D=0$ and the results converge to the exact results rapidly. 
It seems more challenging to deduce the large $n$ behaviors at $D>0$. 

Since the seminal work on the numerical conformal bootstrap \cite{Rattazzi:2008pe}, 
positive semidefiniteness has been used to derive bounds on the solutions to 
the DS equations for random matrix models \cite{Lin:2020mme,Kazakov:2021lel}. 
As discussed explicitly in \cite{Lin:2020mme}, there exist nearly null eigenvectors on the boundary of the non-negative region, 
corresponding to vanishing eigenvalues. 
This is reminiscent of the vanishing Kac determinant \cite{Kac} in the $D=2$ minimal model conformal field theories (CFTs) \cite{Belavin:1984vu}. 
In fact, the majority of the minimal models are non-unitary, 
such as the well-known $\mathcal M(5,2)$ minimal model associated with the Yang-Lee edge singularity \cite{Cardy:1985yy}. 
Inspired by the fact that the minimal models are characterized by the existence of many null states, 
it was proposed in \cite{Li:2022prn} that one can bootstrap non-conformal quantum systems by the null state condition, which is applicable to both unitary and non-unitary theories. 
Along these lines, we propose a novel approach to resolve the indeterminacy of the DS equations 
based on null states in this Letter. 
\\

\paragraph{Dyson-Schwinger equations.}
Before discussing our proposal and concrete examples, 
let's give a brief overview of the DS equations. 
The Green's functions of single scalar field are
\be
G_n(x_1,\dots,x_n)\equiv
\<T\{\phi(x_1)\dots\phi(x_n)\}\>\,,
\ee
where $T\{\dots\}$ indicates that the operators are time ordered. 
In the expectation value $\<\dots\>=\<\O|\dots|\O\>$, 
the state $|\O\>$ is usually assumed to be a ground state. 
The normalization is set by $\<1\>=1$. 
The Green's functions can be obtained from the generating functional $Z[J]$ 
by taking functional derivatives
\be
G_n(x_1,\dots,x_n)=\frac 1 {Z[0]}\frac{\d^n Z[J]}{\d J(x_1)\dots\d J(x_n)}\Big|_{J\rightarrow 0}\,,
\ee
where $Z[J]$ is defined as a functional integral
\be
Z[J]=\int \mathcal D \phi\, 
e^{-S[\phi]+\int d^Dx J(x)\phi(x)}\,,
\ee
and $S[\phi]=\int d^Dx\,\mathcal L[\phi(x)]$ is the Euclidean action. 
In this work, $G_n$ denotes the full Green's function, 
instead of the connected Green's function from the derivatives of $\log(Z[J])$. 
An infinitesimal change in the integration variable at $x$ leads to a quantum equation of motion
\be
\<{\d S[\phi]}/{\d\phi(x)}\>=\<J(x)\>\,.
\ee
The infinite set of DS equations can be derived from functional derivatives with respective to the classical source $J$ and setting $J$ to zero.
\\

\paragraph{Null state condition.}
Null states should be orthogonal to arbitrary test states. 
In practice, we consider approximate null states 
that are orthogonal to a subset of test states 
\be
\<\text{test}^{({L})}|\text{null}^{({K})}\>
=\< \mathcal O_{\text{test}}^{(L)} \mathcal O_{\text{null}}^{(K)}\>=0\,.
\label{null-condition}
\ee
The null and test states are generated by the action of null and test operators on $|\O\>$.  
The superscripts $K$ and $L$ label the numbers of basis operators involved. 
For a given $K$, a well-chosen $L$ can lead to a determined system \cite{fn1}. 
In unitary theories, we use additional null constraints to select out a discrete set of points on 
the boundary of the non-negative region.  

Below, we show that the solutions to the DS equations with null state constraints 
converge rapidly to the exact values. 
They include the complex solutions, 
such as the $\mathcal {PT}$ symmetric ones in non-Hermitian theories 
\cite{Bender:2022eze,Bender:1998ke, Bender:1999ek, Bender:2007nj,Bender:2010hf,r5}, 
where $\mathcal {PT}$ denotes space \cite{parity} and time reversal. 
Nonperturbatively, these complex solutions cannot be derived from positivity constraints. 
\\

\paragraph{Zero-dimensional theories.}
The $D=0$ generating functional is given by a more standard integral. 
The integration path is associated with the choice of Stokes sectors. 
A simple choice in Hermitian theories is along the real axis. 
Since there is no time coordinate at $D=0$, 
the Lagrangian $\mathcal L$ has no kinetic term and we can only study the $n$-point Green's function at equal time. 
The DS equations associated with the Lagrangian 
$\mathcal L=\frac{g}{n}\phi^n$ are
\be
g\, G_{n+k}=(k+1)\,G_{k}\,,\quad
k=-1,0,1,\dots.
\ee
The general solution reads
\be
G_{mn+k}=\left(\frac n g\right)^m \left(\frac {k+1}n\right)_m G_k\,,\quad 
G_{n-1}=0\,.
\label{sol-0D}
\ee
where $k=0,1,\dots,n-1$. 
We have $G_0=1$ by definition, so there are $n-2$ free parameters. 
For $n>2$, indeterminacy is an intrinsic feature of the DS equations, 
rather than a consequence of the finite truncation \cite{fn2}.

Let us first consider the Hermitian quartic theory with  
$\mathcal L=\frac 1 4 \phi^4$. 
Assuming that parity symmetry is not broken, we have $G_{1}=0$, 
so there is only one free parameter, i.e. 
the two-point Green's function $G_2$.  
The exact value of $G_2$ is  
\be
G_2=\pm\frac{ 2 \G(3/4)}{\G(1/4)}=\pm  0.675\, 978\, 240...\,,
\label{phi4-G2}
\ee
where the plus(minus) sign is associated with an integration path along the real(imaginary) axis. 

To solve for $G_2$, we impose the null state condition \eqref{null-condition} with 
\be
\mathcal O_{\text{null}}^{(K)}=\sum_{m=0}^K a_m \phi^m\,,\quad
\mathcal O_{\text{test}}^{(L)}=\sum_{m=0}^L b_m\phi^m\,.
\label{n-t-operators}
\ee
The null state condition should be valid for arbitrary $b_m$, 
so we have $(L+1)$ equations. 
The number of free parameters in the null operator is $(K+1)$. 
After fixing the normalization, 
there remain $K$  parameters. 
We choose $L=K$ to obtain a determined system. 

For $K=3$, the solutions for $G_2$ are $\{-1,\pm 1/\sqrt 3\}$. 
The latter pair of solutions are the same as the one from setting the connected part of $4$-point Green's function to zero, 
whose error is $-14.6\%$ compared to the exact value \cite{Bender:2022eze}. 
As shown in \cite{Bender:2022eze}, the results do not converge to the exact values by setting the connected part of higher-point functions to zero. 
In contrast, our higher $K$ results from the null state condition \eqref{null-condition} converge to the exact values rapidly. 
 
For $K=5$, the solutions are $\{-2/3, 0, \pm1/\sqrt 2\}=\{-0.6667\dots, 0, \pm 0.7071\dots\}$. 
The solution at zero seems unphysical, 
while the errors in the other solutions are $\{-1.4\%, 4.6\%\}$. 
For $K=7$, the solutions are $\{\pm0.5387\dots, \pm 0.6752\dots,  \pm 0.6787\dots,  \pm 1.098\dots\}$. 
The errors in the second and third pairs are $\{-0.12\%, 0.40\%\}$. 
For $K=9$, there is one solution at zero and four pairs of solutions with errors 
$\{-0.0096\%, 0.034\%, -2.5\%, 7.5\%\}$. 
For $K=11$, we find six pairs of solutions and their errors are 
$\{-0.00076\%, 0.0027\%, -0.26\%, 0.82\%, -24\%, 71\%\}$. 
The solutions for the null operators exhibit definite parity.

In general, the solutions for $G_2$ correspond to the roots of some high-degree polynomial equations. 
Most of the roots accumulate around the exact values and 
the errors in the most accurate roots decrease rapidly. 
As the signs of the errors change alternatively, 
we can extract an accurate estimate from the median of a set of roots. 
A robust procedure of extracting the median is to repeatedly throw away the root with the largest distance from the average value, which also applies to the case of complex solutions.  

For example, there are ten pairs of solutions at $K=19$. 
Eight of the positive solutions are in the range $(0.67,0.69)$, 
while four of them are in the tiny range $(0.675978,0.675979)$, 
The median gives $0.6759782403\dots$ and reproduces the numerical expression in \eqref{phi4-G2}. 
The rapid growth of the root density 
is reminiscent of the critical behavior of the Yang-Lee edge singularity \cite{Yang:1952be,Lee:1952ig,Kortman:1971zz}, 
described by the $i\phi^3$ theory \cite{Fisher:1978pf}. 

The second example is the non-Hermitian cubic model $\mathcal L=\frac i 3\phi^3$. 
The only free parameter is $G_1$, whose exact value is
\be
G_1=-3^{1/3}\frac{\G(2/3)}{\G(1/3)} e^{\frac{i2k\pi}{3}}i
=-0.729\,011\,133\dots e^{\frac{i2k\pi}{3}}i
\,,\quad
\ee
where $k=0,1,2$ depends on the choice of the Stokes sectors. 
We impose the null state condition as in the quartic case \eqref{n-t-operators} with $L=K$. 
The solutions again exhibit the phenomenon of root accumulation 
near the exact values. 
For $K=2$, the solutions for $G_1$ are $-2^{-1/3}e^{\frac{i2k\pi}{3}}i$.  
The same results can be obtained 
by setting the connected part of $G_3$ to zero \cite{Bender:2022eze}. 
As $K$ increases, the results of the cubic model converge more rapidly than the quartic case. 
At $K=10$, there are eleven solutions with two of them at zero. 
The remaining nine solutions form three groups of roots related by 3-fold symmetry. 
One group consists of purely imaginary solutions around $-0.73i$, 
which is expected to be $\mathcal {PT}$ symmetric.  
The median gives a remarkably accurate estimate $-0.729011134\dots i$. 

In general, the $\mathcal L=\frac g n \phi^n$ theory has $n$-fold symmetry. 
The above quartic case has $2$-fold symmetry due to parity symmetry. 
For the sextic theory with $\mathcal L=\frac 1 6 \phi^6$, parity symmetry implies that 
the free parameters are $G_2$ and $G_4$. 
We should set $L=K+2$ because of parity constraints.  
We find three groups of $G_2$ roots related by 3-fold symmetry. 
For $K=10$, the median of the root group around the real axis gives $G_2=0.57861625\dots$, 
which is close to the exact value $6^{1/3}\sqrt\pi/\G(1/6)=0.57861652\dots$. 

If we consider the quartic theory $\mathcal L=-\frac 1 4\phi^4$ without parity symmetry, 
then $G_1$ does not vanish. 
The null state condition with $L=K+1$ leads to a determined system. 
For $K=10$, there are 66 roots for $G_1$ and 44 of them are located around $i^k$ with $k=0,1,2,3$. 
The median of the $(-i)$ group gives $-0.977741049\dots i$,  
which is fairly close to the exact solution with $\mathcal {PT}$ symmetry at $-2i\sqrt{\pi}/\G(1/4) =-0.977741067\dots i$. 
For the quintic theory with $\mathcal L= -\frac i 5 \phi^5$, 
we need to set $L=K+2$ and there are 10 groups of roots. 
For $K=10$, the two independent groups of roots give $-1.078676\dots i$ and $0.41184\dots i$, corresponding to
the $\mathcal {PT}$ symmetric solutions at $G_1=-1.078653\dots i$ and $G_1=0.41201\dots i$. 
\\

\paragraph{One-dimensional theories.}
As a natural extension of the $D=0$ procedure, 
we first study the equal-time limit of $D=1$ theories, 
then we consider the situation with unequal time. 
In quantum field theory, the coincidence or zero-separation limit of $n$-point Green's functions is also known as 1-point functions of composite operators. 
In our examples, this limit is regular and 
the typical issue of additional divergences at higher $D$ does not appear. 

In the equal-time limit, one should be careful about the order of operators. 
The contact terms on the right hand side of the DS equations imply 
\be
\<\cdots\big[\dot \phi(\t) \phi(\t)-\phi(\t) \dot\phi(\t)+1\big]\cdots\>=0\,,
\label{commutation}
\ee
which is the counterpart of the canonical commutation relation 
in the operator formalism. 
Translation invariance implies
\be
\<\dot{\mathcal O}_1(\t_1){\mathcal O_2}(\t_2)
+{\mathcal O_1}(\t_1)\dot{\mathcal O}_2(\t_2)\>=0\,.
\label{2pt-time-dot}
\ee
If $\mathcal O_2=1$, we have $\frac d {d\t}\<{\mathcal O}_1(\t)\>$=0, 
the counterpart of $\<[H,\mathcal O_1]\>=0$ in the Hamiltonian formalism \cite{Han:2020bkb}. 

Together with the dynamical part on the left hand side of the DS equations, 
we can express the 1-point functions of composite operators in terms of
\be
F_n={\pa_{\t_2}^nG_2(\t_1,\t_2)}\big|_{\t_1\rightarrow\t_2+0^+}=\left\<\phi(\t)\,\frac {d^n\phi(\t)}{d\t^n}\right\>\,,\quad
\ee
which is independent of $\t$ due to translation invariance. 
In general, we may need more 2-point functions 
as some observables are not related to $G_2$ by DS equations. 

To determine the independent parameters in $\{F_n\}$, we impose the null state condition \eqref{null-condition} with
\be
\mathcal O_{\text{null}}^{(K)}=\sum_{m=0}^K a_m \frac {d^m\phi(\t)}{d\t^m}\,,\quad
\mathcal O_{\text{test}}^{(L)}=\sum_{m=0}^L b_m \frac {d^m\phi(\t)}{d\t^m}\,.\quad
\ee
The null constraints can be expressed in terms of $F_n$ using \eqref{2pt-time-dot}. 
The solutions are again associated with some high-degree polynomial equations. 
As in the $D=0$ case, the majority of roots accumulate around the exact values. 
We can also reconstruct other physical observables from the solutions. 
In particular, the polynomial equation associated with the null state encodes the spectral information. 

The first concrete example is again the quartic theory  
$\mathcal L=\frac 1 2(\dot\phi)^2+\frac 1 2 m^2 \phi^2+g\phi^4$. 
We set $m=1$ and $g=1/2$ to make contact with the previous results 
based on the Hamiltonian formalism. 
The DS equations are
\be
&&\left(-\pa^2_{\t_1}+1\right)G_n(\t_1,\t_2,\dots)+2 G_{n+2}(\t_1,\t_1,\t_1,\t_2,\dots)
\nn&=&\sum_{i=2}^n\d(\t_1-\t_i)\,G_{n-2}(\t_2,\dots,\t_{i-1},\t_{i+1},\dots)\,.
\ee
If $\t_1\neq \t_{i\neq 1}$, the contact terms are irrelevant 
\be
\<\cdots\big[\ddot \phi(\t_1)-\phi(\t_1)-2 \phi^3(\t_1)\big]\cdots\>=0\,,
\label{quartic-DS}
\ee
then the smooth equal-time limit leads to constraints on the 1-point functions of composite operators. 
According to the system of \eqref{commutation}, \eqref{2pt-time-dot} and \eqref{quartic-DS}, 
they can be expressed in terms of $F_n$. 
Some examples are
\be
\<\phi\dot\phi\>=\frac 1 2\,,\quad
\<(\dot\phi)^2\>=-F_2\,,\quad
\<\phi^4\>=-\frac {F_0} 2 +\frac {F_2} 2 \,,\quad
\ee
\be
\<\phi^3\dot\phi\>=\frac {3F_0} 2 \,,\quad
\<\phi^2(\dot\phi)^2\>=\frac 1 2 +\frac {F_2} 6 -\frac {F_4} 6 \,.
\ee
The non-zero 1-point functions involve even numbers of $\phi$ due to parity symmetry. 
Higher time derivatives are removed by \eqref{quartic-DS}.
The independent composite operators take the ordered form $\phi^m (\dot\phi)^n$ due to \eqref{commutation} \cite{fn3}. 
The odd case $F_{2m+1}$ do not appear because they are not independent, such as
\be
F_1=\frac 1 2\,,\quad
F_3=\frac 1 2 +3 F_0\,,\quad
F_5=\frac 1 2-3F_0+9F_2\,.\quad
\ee
Therefore, all the 1-point functions of composite operators are encoded in 
the equal-time limit of $G_2(\t_1,\t_2)$. 
We set $L=2K$ to obtain a determined system. 

Let us examine the simplest case $K=1$ with $a_0\neq 0$. 
The constraints from $\{1,\pa_{\t_1},\pa_{\t_1}^2\}\<\phi(\t_1)\mathcal O^{(K)}_\text{null}(\t_2)\>|_{\t_1\rightarrow \t_2}=0$ are
\be
\left\{\frac {a_1}{2a_0}+F_0\,,\,\frac 1 2 +\frac{a_1}{a_0}F_2\,,\,
\frac {a_1} {2a_0}+\frac{3a_1}{a_0} F_0+F_2\right\}=0\,.
\quad
\ee
This system implies that $F_0=\<\phi^2\>$ is a root of the polynomial $24x^3+4x^2-1$.  
The real root at $0.2991\dots$ is already close to the exact value $\<\phi^2\>=0.3058136507\dots$. 
We may further require that the null state condition is satisfied beyond the equal-time limit, 
then we obtain a differential equation for the 2-point function with $\t_1>\t_2$
\be
(a_0+a_1\pa_{\t_2})\,G^{(K=1)}_2(\t_1,\t_2)=0\,,
\ee 
whose solution is 
\be
G_2^{(K=1)}(\t_1,\t_2)=c_1 e^{\frac{a_0}{a_1}|\t_1-\t_2|}\,.
\ee
The real root of $F_0$ corresponds to $-a_0/a_1=1.6717\dots$. 
This is close to the exact energy gap $E_\text{gap}=E_1-E_0=1.628230531\dots$ 
in the Hamiltonian formulation. 

At higher $K$, the number of solutions also increases. 
The null state condition on the 2-point function implies
\be
G_2^{(K)}(\t_1,\t_2)= \sum_{m=1}^K c_m\, e^{-\D E_m|\t_1-\t_2|}\,,
\label{sol-2pt}
\ee
where $\{\D E_m\}$ are the roots of the null polynomial 
\be
\sum_{m=0}^K a_m\, x^m\,
\label{null-polynomial}
\ee
from the solution for the null operator. 
The interpretation of $\{\D E_m\}$ is the energy differences of intermediate states and $|\O\>$. 
If the energy spectrum is real and $|\O\>$ is the lowest energy state $|0\>$, 
then the roots of \eqref{null-polynomial} should be positive for a gapped system. 
For each $K$ examined, it turns out that there is only one solution with all roots positive! 
In this way, we can make a definite selection from 
a large number of solutions.

\begin{figure}[h!]
\begin{center}
\includegraphics[width=8cm]{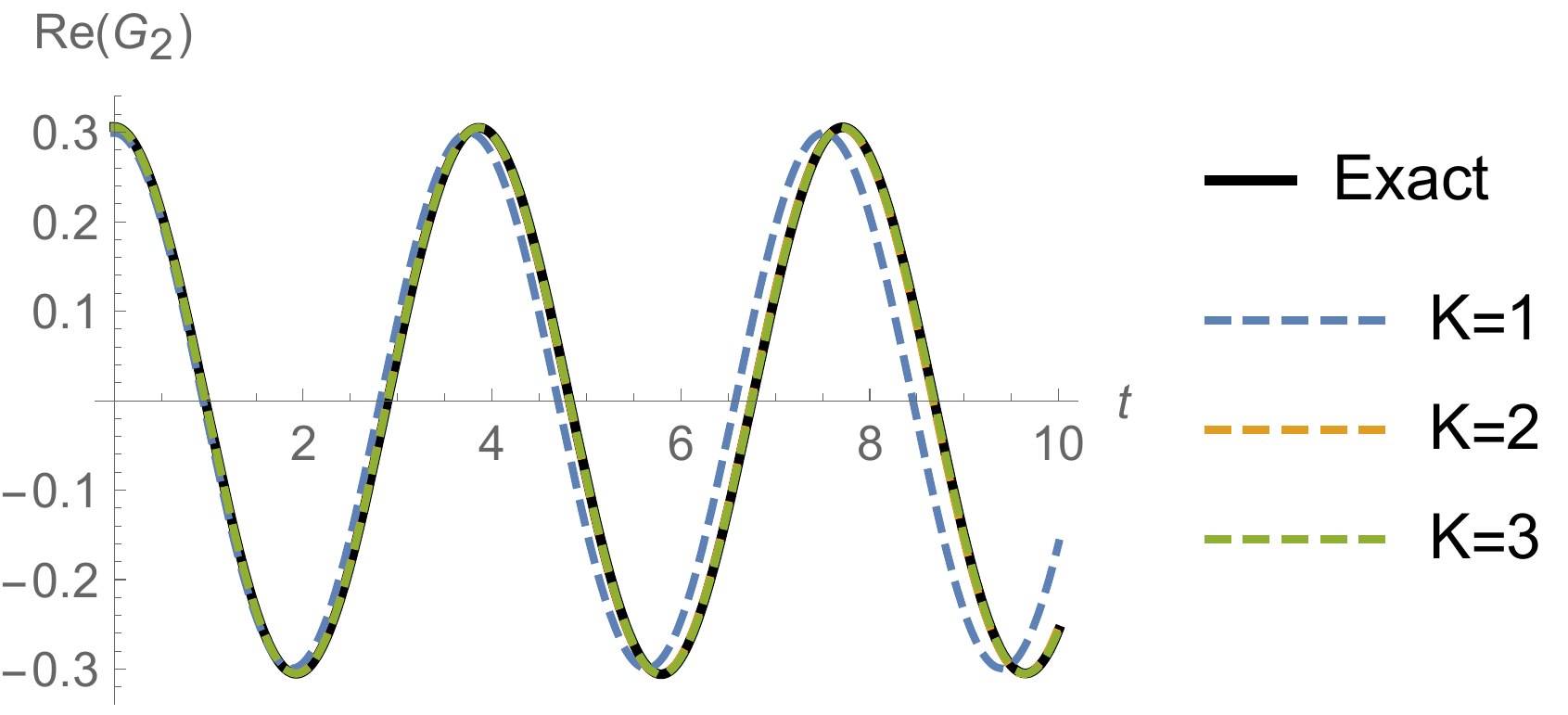}
\caption{The real part of the real-time 2-point function of the quartic theory for $K=1,2,3$ and the exact function. } 
\label{figure-2pt-Re}
\end{center}
\end{figure}

The results converge rapidly to the exact values as $K$ increases. 
In Fig.\ref{figure-2pt-Re}, we present the real part of the two-point function as a function of real time for $K=1,2,3$, where $t=-i(\t_1-\t_2)$. 
The imaginary part also converges to the exact function rapidly. 
For $K=6$, we obtain 
$\<\phi^2\>=0.305813644\dots$ and 
$\D E = \{ 1.628230589\dots ,$ $5.882239\dots, 10.9536\dots , 16.661\dots, 23.3\dots, 32.5\dots\}$. 
For comparison, the exact energy differences between the ground state and 
the low-lying states with odd parity are
$\{1.628230531\dots, 5.882226\dots, 10.9525\dots, 16.624\dots,$ $ 22.8 \dots, 29.4 \dots \}$. 
The estimates are fairly accurate for the low-lying states and reasonably good for higher states. 
We can further determine $c_m$ in \eqref{sol-2pt} using $F_n$, 
which are related to the matrix element $\<n|\phi|0\>$. 
For instance, the square roots of the leading coefficients $c_1^{1/2}=0.5525659561\dots$ and 
$c_2^{1/2}=0.021994704\dots$ are close to the exact values
$|\<1|\phi|0\>|=0.5525659593\dots$ and 
$|\<3|\phi|0\>|=0.021994761\dots$.

The $D=1$ exact solution on the real axis is also a root accumulation point. 
If we do not use the spectral constraint, 
the median of a dense group of roots around the real axis also leads to the same result or 
a small set of nearby roots. 
In the complex plane, there exists a conjugate pair of root groups around $-0.255\pm 0.297i$.  

The second example is the non-Hermitian cubic theory 
$\mathcal L=\frac 1 2(\dot\phi)^2+g\phi^3$ with $g=i/2$.
The DS equations are
\be
&&-\pa^2_{\t_1}G_n(\t_1,\t_2,\dots)+\frac {3i} 2 G_{n+1}(\t_1,\t_1,\t_2,\dots)
\nn&=&\sum_{i=2}^n\d(\t_1-\t_i)\,G_{n-2}(\t_2,\dots,\t_{i-1},\t_{i+1},\dots)\,.
\label{cubic-DS}
\ee
The system of \eqref{commutation}, \eqref{2pt-time-dot}, \eqref{cubic-DS} 
determines the 1-point functions of composite operators and some $F_n$, such as
\be
\<\phi\>=-\frac {2i} 3F_3\,,\quad
\<(\dot\phi)^2\>=-F_2\,,\quad
\<\phi(\dot\phi)^2\>=\frac {i} 3 F_4\,,\quad
\ee
\be
F_0=F_5=0\,,\quad
F_1=\frac 1 2\,,\quad
F_6=-\frac {15}4\,,\quad
F_7=-\frac{45}2F_2\,.\nn
\ee
Then we use the null state condition as in the quartic case. 
Since $\<\phi\>$ does not vanish, 
the null operator should contain a constant term $-a_0\<\phi\>$ due to 
$\<\mathcal O^{(K)}_\text{null}\>=0$. 
We again set $L=2K$ and 
the simplest approximation $K=1$ gives 
\be
\left\{\frac {a_1}{2a_0}-\<\phi\>^2\,,\,
\frac 1 2 +\frac{a_1}{a_0} F_2\,,\,
F_2+\frac{a_1}{a_0} F_3\right\}=0\,.
\ee
This leads to a degree-5 polynomial equation $(F_3)^5=\frac{81}{128}$ with $F_2=\frac 9 {16}F_3^{-2}$, 
so the solutions of $\<\phi\>$ exhibit 5-fold symmetry. 
The real solution at $\<\phi\>=-12^{-1/5}i=-0.608\dots i$ is close to the exact value $-0.590072533\dots i$. 
The estimate of the energy gap $-a_0/a_1=(9/2)^{1/5}=1.351\dots$ is the same as a truncation result in \cite{Bender:1999ek},  
which is not far from the exact value $1.4764808747\dots$. 
At higher $K$, the roots also exhibit 5-fold symmetry and five accumulation points. 

\begin{figure}[h!]
\begin{center}
\includegraphics[width=8cm]{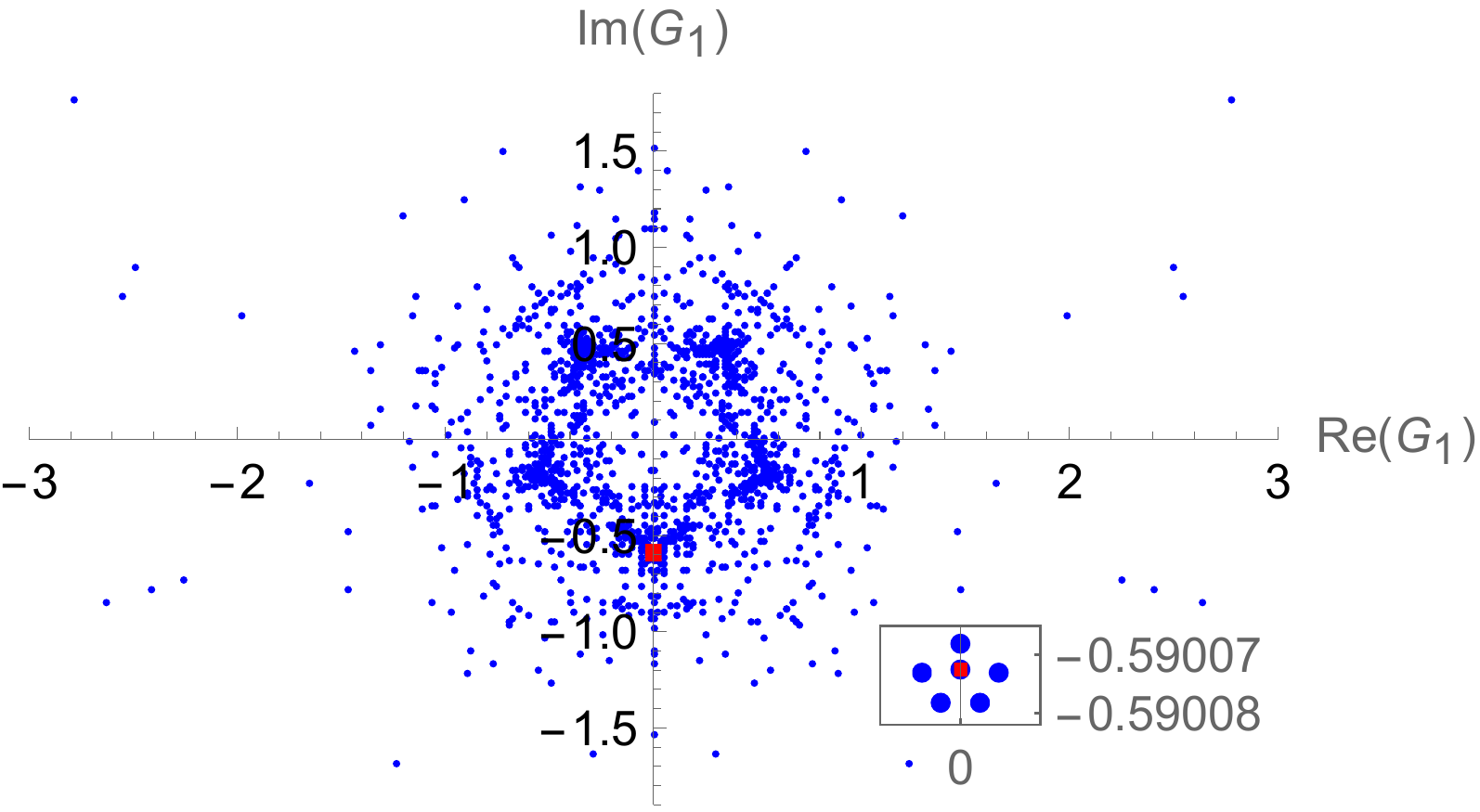}
\caption{The $K=6$ solutions for the 1D non-Hermitian $i\phi^3$ theory.  
The red square indicates the exact value at $G_1=-0.5900725\dots i$.  
We find 123 roots of distance less than $10^{-1}$ from this exact value, 
while $\{44,24,12,6\}$ of them are of distance less than $\{10^{-2},10^{-3},10^{-4},10^{-5}\}$. 
Inset: The 6 solutions are obtained by iteratively discarding the most distant root from the average. 
  } 
\label{figure-root-cubic}
\end{center}
\end{figure}

For  $K=6$, the $G_1$ roots are presented in Fig. \ref{figure-root-cubic}. 
We obtain $\<\phi\>=-0.590072522\dots i$ by assuming the real part of the roots of the null polynomial are positive and the lowest one is has no imaginary part. 
The resulting energy differences \{1.4764812\dots, 3.202970\dots, 5.081\dots,7.008\dots, 9.45\dots$\pm$ 1.05\dots i\} 
are close to the exact values \{1.4764809\dots, 3.202996\dots, 5.079\dots, 
7.059\dots, 9.16\dots, 10.09\dots\}. 
Note that the last two real values become a conjugate pair in the approximate solution. 
As shown in Fig. \ref{figure-root-cubic}, 
the median of the root group around $\<\phi\>\approx-0.6i$ gives the same result. 
In the end, we can determine $c_m$ using $F_n$, such as 
$c_1=0.35645151\dots$ and $c_2=-0.008363555\dots$. 
The exact values are 
$|\<1|\phi|0\>|^2=0.35645139\dots$ and $|\<2|\phi|0\>|^2=0.008363569\dots$. 
\\

\paragraph{Summary.}
We have shown that the indeterminacy of the DS equations can be successfully resolved by the null state condition. 
In some sense, the null state condition can be viewed as a quantization condition, 
playing a similar role as the boundary condition on a functional integral.  
We discovered that the exact solutions are root accumulation points, 
which is analogous to the concentrations of roots in \cite{Bender:2022eze}.  
For the $g\phi^n$ theory at $D=0,1$, 
we obtained rapidly convergent estimates 
from the medians of nearby roots, including the complex solutions,
and reconstructed the time-dependent 2-point Green's function. 
The extensions to higher point functions \cite{Guo:2023gfi} and higher $D$ are in progress.
For more systematic investigations and improved numerical stability, 
it may be useful to apply the tools of computational and numerical algebraic geometry 
or other advanced techniques to solve the polynomial systems. 

At $D>0$, we proposed another method for extracting the best estimate. 
The null polynomial from the null state solution encodes the spectral information 
of intermediate states, 
so a bounded-from-below spectrum should have only positive roots. 
This requirement selects a unique solution. 
The close relations among null states, differential equations for $n$-point functions, 
and intermediate spectra are in beautiful parallel with 
the classical work of Belavin-Polyakov-Zamolodchikov 
on 2D minimal model conformal field theories \cite{Belavin:1984vu}. 

We elucidated the intimate connection between 
approximate external null states and truncated spectra of intermediate states. 
It would be interesting to revisit the general $D$ conformal bootstrap methods \cite{Poland:2018epd}.  
The truncation approach initiated by Gliozzi \cite{Gliozzi:2013ysa} 
can be viewed as a general $D$ approach based on approximate null states. 
The present Letter provides new insights on how to select the best estimates from 
the solutions to truncated crossing equations. 
Furthermore, the complex solutions in the CFT context can have important implications on 
gauge theory, statistical and condensed matter physics \cite{Gorbenko:2018ncu}. 
\\

I would like to thank the referees for the constructive comments and suggestions. 
This work was supported by the 100 Talents Program of Sun Yat-sen University  
and the Natural Science Foundation of China (Grant No. 12205386).


\begin{thebibliography}{10}
\bibitem{parity}
Under the parity transformation $\mathcal P$, we have $\phi\rightarrow -\phi$. 
We can view $\phi$ as a space coordinate in quantum mechanics.
\bibitem{fn3}
It may be useful to deduce their asymptotic behavior at large $m,n$.
\end{thebibliography}
\end{document}